\documentclass{eas}
\usepackage{graphicx}

\begin{document}
\title{Deep Field Surveys - A Radio View} 
\author{M.~A.~Garrett}\address{Joint Institute for VLBI in Europe,
  Postbus~2, 7990~AA Dwingeloo, The Netherlands
\email{garrett@jive.nl}} 

%\thanks{The National Radio Astronomy
%Observatory is a facility of the National Science Foundation operated
%under cooperative agreement by Associated Universities, Inc.}
%
\begin{abstract}
  
  I present an introductory review text, focusing on recent
  deep field studies using radio telescopes, including high-resolution,
  wide-field VLBI observations.  The nature of the faint radio source
  population is discussed, taking into account complimentary data and
  results that are now available via the sub-mm, IR, optical and X-ray
  wave-bands.  New developments regarding the increased sensitivity of
  VLBI and an expansion in the instruments field-of-view are also
  presented. VLBI may be an important tool in recognising distant,
  obscured AGN from the ``contaminant'' star forming galaxies that now
  dominate lower-resolution, sub-arcsecond and arcsecond radio
  observations.  

\end{abstract}
\maketitle
\section{Introduction - How many radio sources are in the sky ?}

%3rd of July 1987

On a sweltering July summer's day in 1987, in the midst of the rural
Cheshire plains, Sir Bernard Lovell gave a rousing speech,
commemorating the 30th anniversary of the completion of the MkI radio
telescope at Jodrell Bank (renamed on that day the Lovell Telescope,
see Gunn this volume). The precise date was 3 July 1987, surrounding
the telescope a great Jodrell party was in full-flow, involving several
hundred VIPs, plus a few new (and by late afternoon very happy) Jodrell
Bank PhD students.  In his speech, Sir Bernard made a simple statement,
but it was one that shocked and inspired all at once. Quite simply
he stated:

\begin{center}
{\it There are 4 million radio sources in the Sky!} 
\end{center} 

For some of the students this was something of a revelation, 4 million
was a big number, much more than most of us had expected. At the same
time, we were naive enough to wonder just how Sir Bernard could be so
sure of his facts and figures - had he counted all 4 million sources
himself ?! The size of the number also brought hope - 4 million sources
must surely mean that there was enough interesting objects to share
between even the largest group of aspiring (and as I was beginning to
realise) very competitive students, not to mention satisfying the
criteria for a successful PhD and thesis defence.

But Sir Bernard's statement also raised a few doubts, at least in my
mind. For example, why, if there really were 4 million radio sources out
there, did the discussion during afternoon-tea at Jodrell Bank, always
revolve around the same objects --- all with inexplicable names that
were now becoming all too familiar - 3C~273, SS~433, Cyg-A, the ``Double
Quasar'' and one or two others with long telephone numbers attached.
While I was to embark on a thesis focussed on one of these ``famous
sources'', in this paper I am glad to return to the 4 million others,
Sir Bernard originally inspired us with.

\subsection{Radio Source Counts} 

In the mid-1980s, the Westerbork Synthesis Radio Telescope (e.g.
\cite{win85}, \cite{oor87}) and the Very Large Array (e.g.
\cite{mit85}) had already begun to shed light on the source counts at
the milliJansky and sub-milliJansky flux density levels. By observing
particular and relatively small areas of sky, the sub-mJy source counts
were found to be fairly consistent between different fields, and
were thus well established by the time of the Lovell Telescope's
30th birthday party.

\begin{figure} % Figure 1
\includegraphics[height=8cm,width=12.5cm]{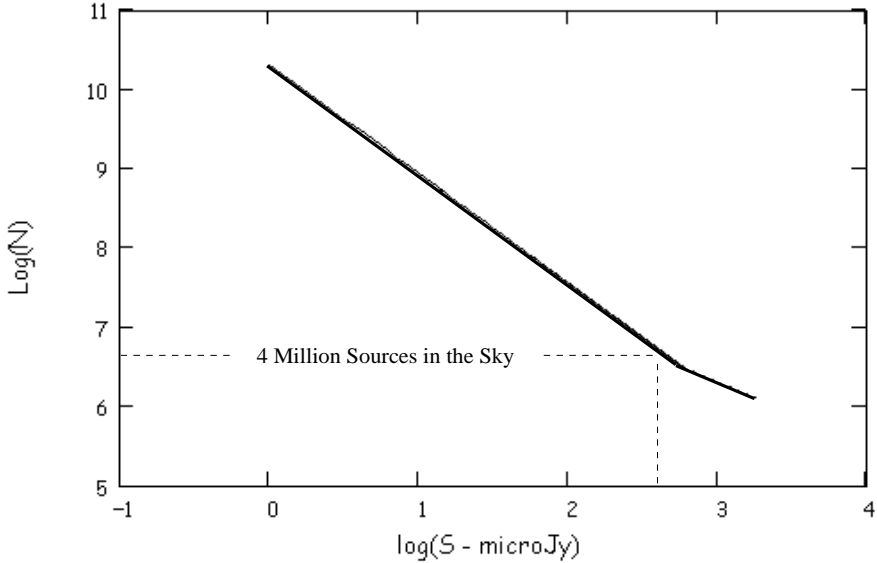}
\caption{The integral radio source counts $(log N)$ - $(log S)$ at 1.4 GHz
  over the mJy, sub-mJy and microJy flux density range. At the 0.5 mJy
  detection threshold, there are about $\sim 4$ million radio sources
  in the sky. At 40 microJy that number increases to $\sim 100$ million
  radio sources.  }
\end{figure}

The integral radio source counts, $N(>S)$, are usually represented as a
power-law fit in $S$, $N(>S)=kS^{-\gamma}$), but often what is plotted are
the {\it differential} radio source counts (e.g. Condon this volume,
Fig.~X). While these are more useful in terms of studies of source
evolution or cosmological tests, the integral counts directly predict
the number of sources we can expect to observe for a given flux density
threshold (telescope sensitivity) and area of sky. Today, recent deep
field radio source surveys (e.g. \cite{ric00}) provide good estimates of
the source count down to a few tens of microJy: 

\begin{equation} 
  N(> S_{1.4 GHz}) \sim  133S^{-1.38} {\rm arcminute}^{-2}     
\end{equation}  

where N is the expected number of sources per square arcminute of sky,
above a specific flux density level, S (in microJansky). 

For an r.m.s. noise level of say $7\mu$Jy, you expect to detect (at the
5-sigma level i.e. $S\sim 35\mu$Jy) about 1 radio source per square
arcminute. Fig.~1 plots this fit of the integral counts over the {\it
  whole sky\/} at 1.4 GHz. As recognised by Sir Bernard in 1987, the
sub-mJy flux densities (then being routinely explored by the best radio
telescopes of the day), correspond to all-sky source counts of several
million radio sources. Note that the slope of the integral source
counts begins to steepen at sub-mJy flux densities ($\propto
S^{-1.38}$), suggesting the emergence of a new population of radio
sources at faint flux density levels. At the typical detection
thresholds routinely probed today ($\sim 40$ microJy), the total
number of sources in the sky thus rises to 100 million radio sources. As
shown pictorially in Fig.~2, at mJy flux density levels the radio sky
is virtually empty, at microJy flux density levels, the radio
sky literally {\it lights-up}. 

\begin{figure} % Figure 2
\includegraphics[height=4.5cm,width=12.5cm]{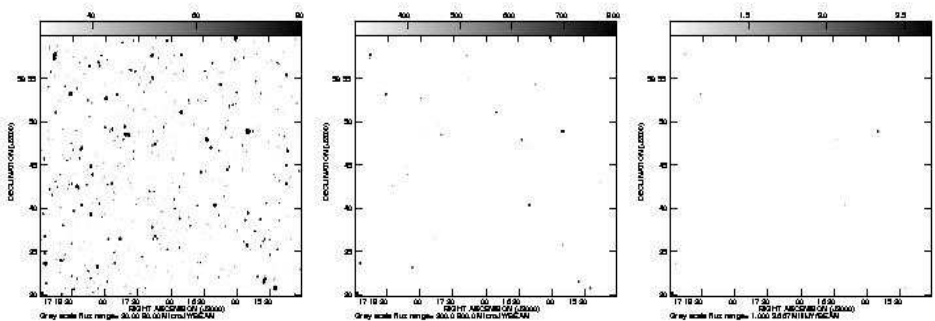}
\caption{An area of radio sky (about the extent of the full moon)
  centred on the Spitzer First Look Survey, as seen by the WSRT at 1.3
  GHz (\cite{mor04}). The region is displayed  
  with various flux density limits - {\it Left}: $S > 30$
  microJy, {\it centre:} $S > 0.3$ mJy and {\it right:} $S > 1$ mJy. 
  At microJy flux density levels, the radio sky literally lights-up. At
  mJy flux density levels the radio sky is virtually empty. 
}
\end{figure}

\section{Nature of the microJy Radio Source Population} 

\subsection{Potential sources of microJy radio emission} 

It is instructive to consider what kind of radio emitting object (or
source population) might be responsible for the up-turn in the faint
microJy radio source counts, as revealed in deep field radio studies.
Fig.~3 plots the observed radio flux density as a function of
increasing redshift for several types of well-studied radio sources
(excluding the very luminous, $L\sim 10^{25-29} W/Hz$, but rare radio
galaxies and quasars that are known to dominate the bright ($>$~mJy)
source counts \cite{con84}. The sources include: 
\begin{itemize} 
\item low-luminosity AGN
as observed in the local Universe (e.g. M84, $L\sim 10^{23}$W/Hz),
\item normal, spiral star forming galaxies like our own Milky Way
  ($L\sim 10^{18-21}$ W/Hz) and nearby starburst systems such as M82 \&
  Arp 220 (in which the radio emission is generated via star formation
  processes, $L\sim 10^{22-23}$),
\item very luminous but rare and transient GRB radio after-glows
- e.g. GRB030329 ($L\sim 10^{24}$ W/Hz),
\item individual luminous hypernovae
in starburst galaxies such as Arp~220 ($L\sim 10^{21-22}$),
\item Supernovae remnants/event such as Cas-A SN1993J ($L\sim
  10^{17-21}$ W/Hz),
\item radio binary star systems during outburst e.g.  Cyg-X3 ($L\sim
  10^{17}$ W/Hz).
\end{itemize}

Figure 3 shows, that of the commonly known continuum radio emitting
objects, the most likely contributors to the microJy radio source
counts (at moderate redshift, $z < 1$) include AGN (low-luminosity FRI
radio sources) and Ultra-Luminous Infra-Red Galaxies (ULIRG) i.e. very
luminous starburst galaxies, such as Arp 220. In principle, GRBs can be
detected within these microJy samples too, but they are much too rare
and short-lived to contribute in any substantial way. Moderate
starburst galaxies, such as M82, can be detected at moderate redshifts,
out to $z\sim 0.4$. The largest, ``normal'' star-forming galaxies (e.g.
galaxies similar to our own Milky Way) might contribute to the source
counts, but at current sensitivity levels they can only be detected
over a limited region of redshift space ($z < 0.2$). It is interesting
to note that luminous SNR (similar to the brightest examples detected
in Arp 220, \cite{smi98}) can be detected individually but again over a
very restricted redshift range ($z < 0.1$). Galactic objects such as
Cas-A, SS433 and Cyg-X3 (in outburst) are not luminous enough to be
detected in current deep radio surveys. However, next generation radio
instruments (such as the SKA - see section \ref{SKA}), should be able
to detect systems like Cyg-X3 (in outburst) out to significant
distances ($z < 0.2$). Dwarf irregular galaxies are considered to be
the most common type of galaxy in the universe, but with radio
luminosities comparable to Cas-A, they are unlikely to dominate the
microJy source counts.

\begin{figure} % Figure 3
\includegraphics[height=6.5cm,width=12.5cm]{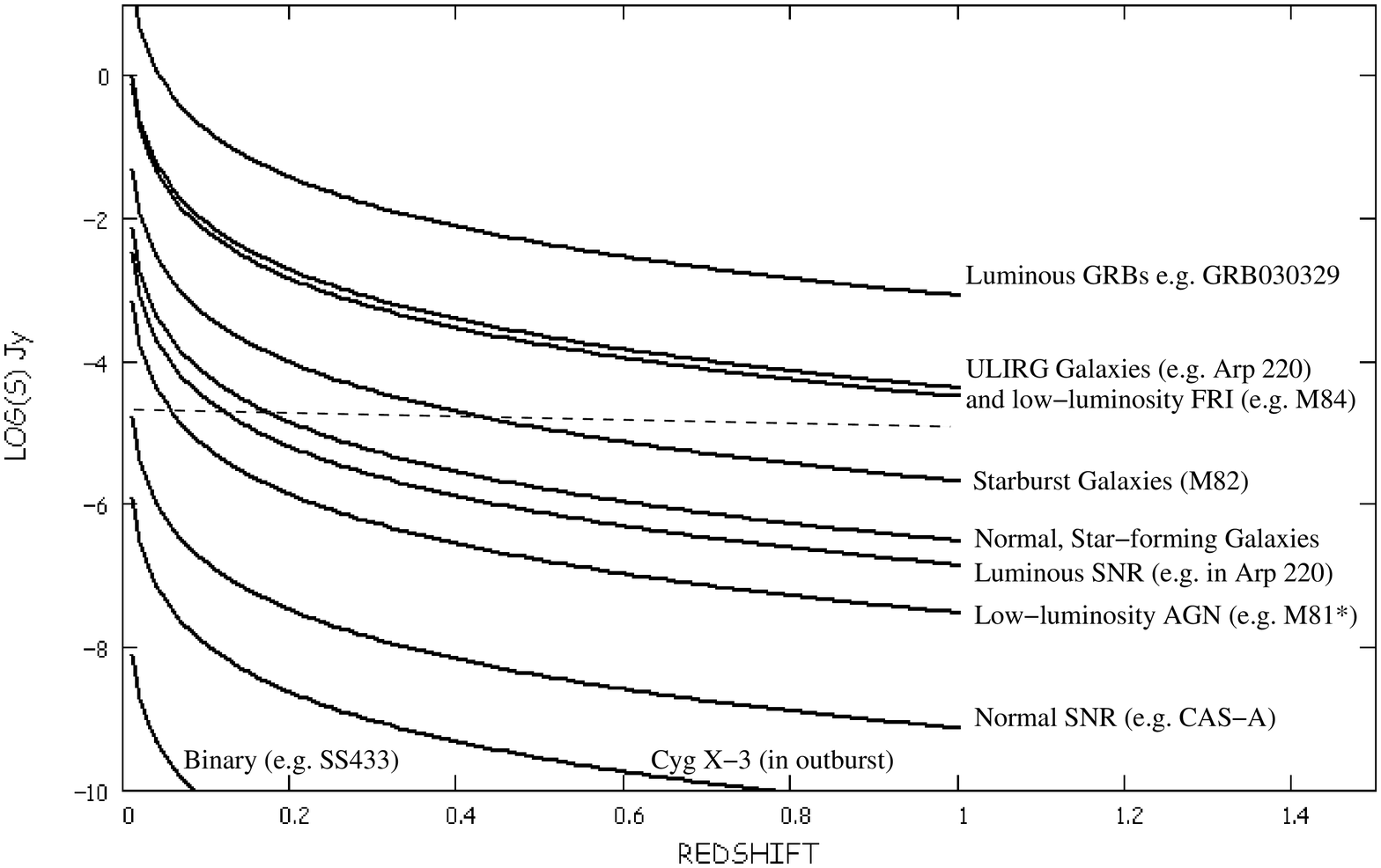}
\caption{A plot of flux density versus redshift for various radio
  source type. The dotted line represents the limiting detection
  threshold of deep radio surveys conducted at 1.4 GHz.}
\end{figure}

\subsection{Optical \& Infra-Red counterparts to faint radio sources in the
  HDF-N} 

Significant progress has been made over the last 5 years in identifying
optical, Infra-red, sub-mm and X-ray counterparts to the faint microJy
radio sources detected in the Hubble Deep Field North (HDF-N,
\cite{wil96}). The great advantage of the original HST image of the
HDF-N, was not just its depth ($I < 29^{m}$) but the excellent angular
resolution achieved. For the first time it became possible to discern
the optical morphology of relatively distant galaxies.  On inspection
it immediately became clear that the most distant galaxies in the HDF-N
were also the most morphologically disturbed.  For example, familiar
``grand design'' spirals observed locally, all but disappear beyond $z
\sim 0.3$. In terms of morphology, these distant, disturbed systems are
most akin to nearby Ultra Luminous Infrared Galaxies (ULIG) and
interacting starburst systems (\cite{abr01}).

While the brightest radio sources in the sky are usually associated
with stellar identifications (quasars), as one moves to fainter flux
density thresholds, an increasing fraction of the identifications are
with galaxies, including bright elliptical galaxies with red colours.
Indeed at the mJy level, around half of the identifications are with
galaxies (\cite{mcm01}). The microJy radio source population continues
this trend -- in the HDF-N the optical counterparts are usually (but
not always - see section~\ref{ofr}), identified with galaxies brighter
than $I = 25^{m}$ (\cite{ric98, fom02, mux04}).

\begin{figure} % Figure 4
\includegraphics[height=5.5cm,width=12.5cm]{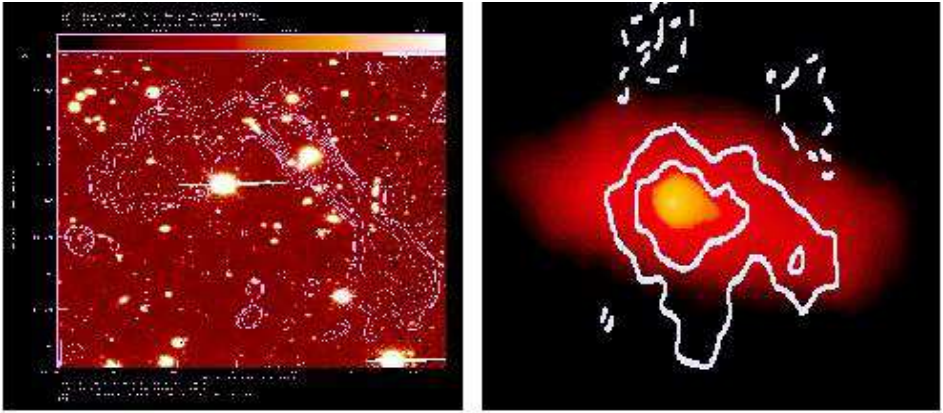}
\caption{WSRT (\cite{mor04}) and
  MERLIN-VLA (\cite{mux04}) radio contour maps of 2 radio sources (a
  bright radio-loud AGN - {\it left} and a faint microJy star forming
  galaxy - {\it right}). The contour maps are projected on top of their
  respective optical backgrounds. The AGN radio emission spans 5
  arcminute. By comparison the radio emission associated with the
  star-forming galaxy is sub-galactic in scale, spanning only a few
  arcsecs.  }
\end{figure}

\subsubsection{Radio morphology, angular size} 
\label{size} 

The galaxies associated with the microJy radio source population are
usually identified with major disk galaxies (rather than elliptical
galaxies), and the radio emission is often sub-galactic in size, i.e.
the scale of the radio emission usually fits within the entire optical
extent of the galaxy (\cite{mux04}). This can also be seen in the WSRT
1.4 GHz deep field observations presented in Fig.~2 - the vast majority
of the sources are unresolved at 15 arcsecond resolution. The VLA also
does not fully resolve these faint radio sources (\cite{ric00}),
however, combined MERLIN-VLA 1.4~GHz observations are able to measured
sizes for the radio sources typically in the range of less than a few
arcseconds (e.g.  Muxlow et al. 2004). This should be compared with the
median angular size of $\sim 10$ arcseconds for the brighter, mJy radio
source population (\cite{col85}). Figure 4 presents WSRT and MERLIN-VLA
radio contour maps of 2 radio sources projected on top of their
respective optical backgrounds. The bright ($S \sim 40$~mJy) radio-loud
AGN extends across the optical field, spanning some 5 arcminutes
(\cite{mor04}). In this case the extent of the radio source is so
large, it is quite difficult to guess which of the galaxies in the
field can be identified with it. The much fainter microJy radio source
($S \sim 40$~$\mu$Jy) is only a few arcseconds across, with most of the
radio emission lying within the optical emission associated with host
galaxy. Note the scales in both images are very different, as are the
radio source sizes.  In terms of detailed morphology (\cite{mux04}),
there are very few (if any), microJy radio sources that exhibit the
familiar and often spectacular ``classical'' radio structures
associated with bright radio galaxies and quasars (e.g.  widely
separated steep spectrum radio lobes straddling an unresolved, usually
fainter, flat spectrum central radio core).

\begin{figure} % Figure 4
\includegraphics[height=5.5cm,width=12.5cm]{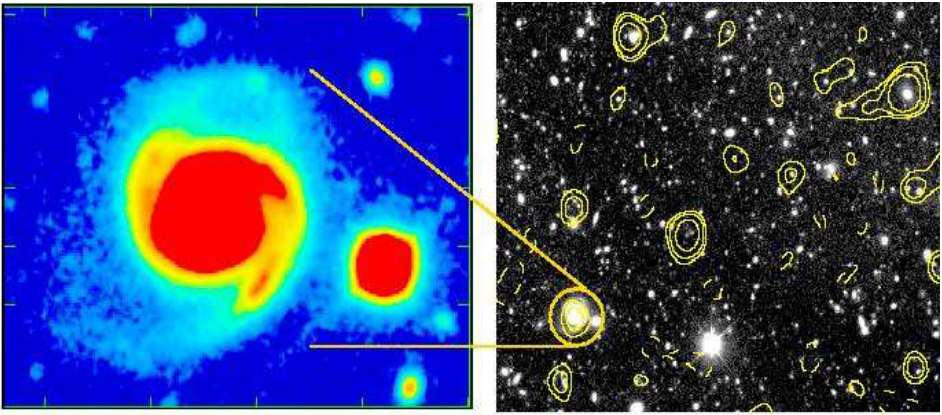}
\caption{A faint but extended microJy radio source (contours {\it
    right}) detected by the WSRT (\cite{gar00}) but not by the VLA or
  VLA-MERLIN combination (\cite{mux04}). Optical observations ({\it
    left}) identify the optical host as a local ($z=0.106$) spiral
  galaxy. In this case, even moderate resolution ($\sim 1-2$
  arcseconds) observations, can potentially resolve-out the radio
  emission from the extended galaxy disk. }
\end{figure}

Nevertheless, it should be noted that the VLA and MERLIN observations
(e.g. Muxlow et al. 2004) probably miss some ($\sim 10$\%) of the
extended sources detected by the lower-resolution WSRT observations
(\cite{gar01}) - these probably include some extended AGN but also
local (and therefore extended) star-forming disk galaxies. Fig.~5 shows
the optical identification (\cite{B98}) of a faint but extended radio
source ($S\sim 67\mu$Jy) that is detected by the WSRT but not by the
VLA or VLA-MERLIN.  The galaxy shows clear spiral structure in the
optical and is located nearby (z=0.106), corresponding to a distance of
$\sim 400$~Mpc. At this distance, the typical optical extent of a major
disk galaxy is $\sim 20$~kpc, thus subtending an angle of $\sim 10$
arcseconds on the sky. The detection of this system by the WSRT (and
non-detection by the VLA or VLA-MERLIN combination) suggests that in
this case the radio emission covers a large fraction of the optical
disk.  While high resolution is essential for uncovering the radio
morphology of distant galaxies, it is wise to remember that even
moderate resolution observations ($\sim 1-2$ arcseconds), can
potentially resolve-out the radio emission from local galaxies.  Deep
surveys therefore require observations that span a range of resolution,
if they are to be considered complete.

\subsection{Redshift Distribution} 

Fortunately, the HDF-N is also well served in terms of spectroscopic
data (e.g. \cite{coh00}), so for most radio sources in the HDF-N, a
redshift is also known. Here again there is a significant difference
between the redshift distribution of the brighter mJy and microJy
source populations. The former are typically located at cosmological
distances ($z \sim 1$), whereas the microJy source population show a
fairly continuous distribution of moderate redshifts, from $z \sim 0.2
- 1.3$

\subsubsection{The FIR-Radio Correlation at high-z} 

Another important clue to the nature of the bulk of the microJy radio
source population is the strong correspondence between the Mid-IR ISO
detections (Aussel et al. 1999) and the WSRT radio detections
(\cite{gar02}). The obvious conclusion is that the FIR-radio
correlation (well established for nearby star forming galaxies e.g.
\cite{Hel93}) also applies at cosmological distances. Fig.~6 shows a
logarithmic plot of the FIR vs Radio Luminosity for local galaxies
(\cite{con92}) presented together with addition points (large circles)
from the (generally) higher luminosity/higher redshift ISO/WSRT HDF-N
sample. After applying appropriate k-corrections to both the
Mid-IR and radio data, the majority of radio sources appear to closely
follow the FIR-radio correlation out to at least $z \sim 1.3$
(\cite{gar02}).

The fact that the bulk of the microJy source population follow the
FIR-radio correlation, together with their optical identification and
steep spectrum radio morphologies, strongly suggests that for most
systems the radio emission is generated by star formation processes,
rather than the accretion of matter onto super-massive black holes
(AGN). It should be stressed that star formation only begins to
dominate at the faintest levels probed by deep surveys ($<
100$~$\mu$Jy), indeed even at the sub-mJy level $(100-1000$~$\mu$~Jy)
AGN still compose a significant fraction of the source counts. However,
at the faintest levels probed by deep surveys ($\sim 40$ microJy at 1.4
GHz), star formation processes (star forming galaxies) appear to
account for $\sim 2/3$ of all sources of radio emission.

In all these discussions of AGN vs Starburst classification, it should
be made clear that we are attempting to characterise the nature of the
dominant process that gives rise to the faint radio emission in sub-mJy
and microJy radio sources. Hybrid systems, presenting both AGN and
star forming characteristics are possible. For example, since AGN are
usually radio quiet, it's quite possible that in any individual galaxy,
radio emission being generated by star formation may also be
accompanied by emission at other wave-bands (e.g. X-rays) being
generated by accretion associated with an (embedded) AGN. It's also
possible that the radio emission is generated by both star formation
and AGN processes simultaneously. The fact that the FIR-radio
correlation appears to apply to so many of the sub-mJy and microJy radio
source population, however, suggests that star formation is the
dominant process by which radio emission arises in these faint radio
sources. 

\begin{figure} % Figure 5
\includegraphics[height=10.0cm,width=10.5cm]{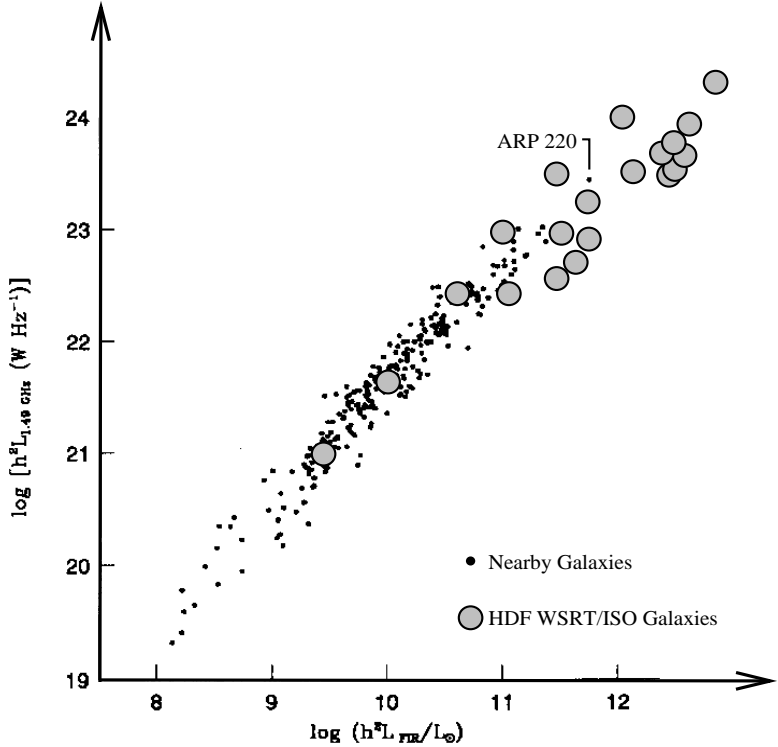}
\caption{A logarithmic plot of the FIR vs Radio Luminosity 
  for local galaxies (Condon 1992) with the addition of the ISO/WSRT
  extension to fainter but higher luminosity/higher redshift sources
  shown by the larger filled circles (Garrett 2002). The additional
  ISO/WSRT sample covers a range of redshift, up to $z \sim 1.3$. The
  most luminous sources are also the most distant, with implied SFRs
  an order of magnitude greater than Arp 220.}
\end{figure}

\subsection{Radio and FIR emission processes} 

It should also be noted that the linear relationship between the non-thermal
radio emission and the star formation rate is not understood in detail.
Neither is the remarkable tightness of the FIR-radio correlation, a
relation that now spans over 6 orders of magnitude in radio luminosity!
Even the source of the electrons that produce the bulk of the radio
emission is unclear - the most widely accepted scenario is that
electrons are accelerated to relativistic velocities in supernova
remnants. A typical supernova associated with a massive star, releases
around $10^{44}$ Joule almost instantaneously. The ejecta and the
circumstellar medium in the immediate vicinity of the SNe is swept up
into a supersonic shell ($v \sim 10000$ km/s) of dense circum-stellar
material, causing a shock wave which in turn heats the surrounding
matter with which it collides. The resulting expanding and cooling
structure of ejecta and shock-heated material is known as a supernova
remnant. It is within these SNR shocks that electrons (and other ions
associated with low-energy cosmic rays) are believed to be accelerated
as the supernova remnant expands into the ISM. The favoured model
suggests that magnetic turbulence in the ISM scatters electrons (and
the other ions) back and forth across shocks in the SNR.  The electrons
continually gain energy at each crossing, in a process known as Fermi
acceleration. This picture of diffuse shock electron acceleration is
supported by the detection of non-thermal x-ray emission in the SN 1006
and other supernova (e.g. \cite{koy95}).

\begin{figure} % Figure 6
\includegraphics[height=4.0cm,width=12.4cm]{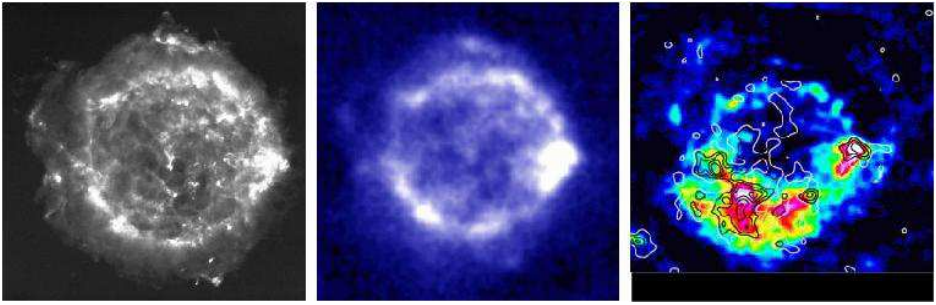}
\vspace{0cm} 
\caption{Radio VLA (left) and sub-mm SCUBA images (middle \& right) 
  of a galactic SNR, Cas-A (\cite{bra87} \& \cite{dun03}). The
  synchrotron-subtracted SCUBA image (right) shows the distribution of
  cold dust in this SNR. Supernovae are one of the key physical
  elements behind many of the observational aspects of star formation -
  in an instant they unleash more energy than the Sun does over its
  entire lifetime. They are chiefly responsible for the radio emission
  seen in star forming galaxies (via particle acceleration) and they
  also give rise to the dust that re-processes the uv radiation from
  young, massive stars (the same stars that rapidly evolve into type II
  Supernovae).}
\end{figure}

The process of electron acceleration lasts well beyond the typical
lifetime over which SNR can usually be detected by current radio
telescopes. Over time scales of many tens of thousands of years, the
electrons eventually find themselves distributed throughout the host
galaxy, radiating synchrotron radio emission as they spiral through the
large-scale galactic magnetic field. Another source of radio emission
is free-free emission associated with thermal bremstraalung in ionised
HII regions, plus prompt and compact synchrotron radio emission from
young SNe and SNR. In most luminous star forming galaxies, the latter
discrete sources only represents about 10\% of the total radio
emission, the remainder being dominated by the distributed cosmic ray
electrons. The exact division between thermal and non-thermal radio
emission is complex, at high frequencies $> 30 $~GHz thermal emission
dominates, at lower frequencies ($< 10$~GHz) synchrotron emission
dominates. It's sobering to note, that the total radio continuum
emission, only accounts for less than $10^{-4}$ of a typical star
forming galaxies bolometric luminosity.

In this scenario, the radio emission is then essentially proportional
to the supernova rate of {\it massive} stars, which in turn is
proportional to the massive star formation rate. Since massive stars
have short lives, the radio emission is a reasonable estimate of the
current, on-going star formation rate. By assuming some reasonable mass
function for stars (e.g. a classical Salpeter Inititial Mass Function,
IMF) the total star formation rate can be estimated by integrating over
the assumed IMF.  While the SFR for massive stars can be determined
reasonably well, estimates of the total star formation rate are less
reliable, in particular the form of the IMF is less well constrained
for low-mass stars.

Both evolved, massive stars, and most recently SNR (e.g. \cite{dun03},
see also Fig. 6) are also thought to be responsible for generating the
cold dust that pervades distant star forming galaxies. This very fine dust is
composed of carbon and silicates, that optimally scatter and absorb
radiation with a wavelength comparable or smaller than the individual
grain size ($\sim 1 \mu$m) i.e.  optical (blue light) and uv radiation.
If the column density of dust is high enough, even x-ray emission can
be absorbed.  This can happen in regions with high star-formation
rates, or in the central region around AGN, where large amounts of dust
also appear to reside.  In addition, for galaxies at high-z, near-IR
observations can also suffer from dust obscuration, since these photons
were emitted in the rest frame of the galaxy at optical wavelengths.
Fortunately at longer wavelengths (e.g. radio, sub-mm, FIR), dusty
star-forming systems and the central regions of AGN, remain
transparent.

The uv-radiation that is emitted by young, massive stars is not only
responsible for producing ionised HII regions, but also gets absorbed
by cold dust, heating the grains in the process. These grains then
re-radiate with a a modified blackbody spectrum ($T \sim 30- 100$~K)
for which the spectral energy distribution (SED) peaks in the FIR at a
few hundred microns.  Just as in the radio, a measure of the FIR
emission is a direct measure of the massive star formation rate. This
partially explains why both the radio and FIR emission are so well
correlated, though the tightness of the relation requires considerable
(perhaps unlikely) fine-tuning of various inter-related physical
parameters. A definitive understanding of this complex correlation is
still awaited. 

\subsection{Sub-mm Galaxies and Optically faint microJy radio sources} 
\label{ofr}

At high redshift ($z > 2$), the peak of the modified blackbody curve
associated with star forming galaxies is shifted (by a factor $1+z$)
from the Far-IR to sub-mm wavelengths. This more than compensates for
the dimming effect of increasing luminosity distance, such that the
apparent luminosity of a dusty star forming galaxy at sub-mm wavelength
remains constant over a large range of redshift $(z \sim 2-10)$. This
so called ``negative k-correction'' has been used to great advantage by
the SCUBA instrument installed on the JCMT in Hawaii \cite{hug98}. Deep
radio observations have been crucial in pin-pointing the location of
the dust enshrouded sub-mm galaxies and identifying their very faint optical
counterparts (e.g. \cite{dun04}).

Another interesting aspect of these studies, is that about 10-20\% of
the faint radio source population remain unidentified to $I=25^{m}$
(\cite{ric99}). This optically faint radio source population seems to
have a significant overlap with the sub-mm SCUBA source population
(\cite{cha01}), suggesting that they are distant, massive galaxies,
obscured by the presence of large amounts of dust.  However, around 1/2
of the optically faint radio sources ($R > 23.5^{m}$) have no sub-mm
counterpart. \cite{cha04} argue that these radio sources represent a
new population of galaxies - an extension of the high-redshift sub-mm
galaxy population, but with hotter characteristic dust temperatures,
shifting the peak of their far-IR emission to shorter wavelengths and
reducing the sub-mm flux below the sensitivity of current instruments
such as SCUBA.  Chapman et al. have concluded that up to 1/2 of the
high-redshift ultra-luminous star forming galaxies are missed by current
sub-mm surveys alone. This is probably an upper-limit, since some of the
optically faint radio sources also show evidence of AGN activity, for
example in their optical spectra. Whether the faint radio emission
arises purely from star formation processes or also (perhaps partially)
from AGN activity is difficult to tell from the limited information
available. As we shall see in the next section, rapid developments in
very high resolution radio astronomy (e.g. VLBI) should be able to
address this question directly.

\section{VLBI Observations of Deep Field Survey Regions -
  distinguishing between Star forming and AGN systems} 

As we have seen in section \ref{size}, the faint microJy radio
source population are resolved at sub-arcsecond resolutions
(\cite{mux04}), limiting the maximum brightness temperature of star
forming galaxies to $< 10^{5}$~K at frequencies above 1 GHz
(\cite{con92}).  High-resolution (milliarcsecond) VLBI observations
resolve out these extended star forming galaxies and, currently, any
compact radio sources in these galaxies, such as luminous Type IIn SNe
(\cite{smi98}), would be too faint to detect at cosmological distances.
On the other hand, VLBI is very well matched to detect the very compact
radio emission associated with the relativistic outflows generated by
accretion onto massive black holes in active galaxies.

High sensitivity VLBI observations, covering a large fraction of sky,
thus provide a simple and direct method of identifying these faint, and
possibly distant, radio-loud active galaxies.  Free from the effects of
dust obscuration, deep and wide-field VLBI studies can therefore
contribute to the cosmic census of active galaxies (and their
energising massive black holes), and together with redshift
information, can potentially probe the accretion history of the early
Universe. In addition, the positions of the compact VLBI detections can
be measured very accurately, via the use of standard phase-referencing
techniques. Astrometric precision at the mas level can be achieved
routinely (e.g. \cite{wro04}), and cross identification with sources detected
at other wave-bands can anchor these non-radio observations to the
International Celestial Reference Frame (ICRF). In crowded and deep
fields, such precision astrometry may be useful in identifying the
counterparts of faint or obscured sources at other wavelengths.

\begin{figure} % Figure 6
\includegraphics[height=12.0cm,width=12.4cm]{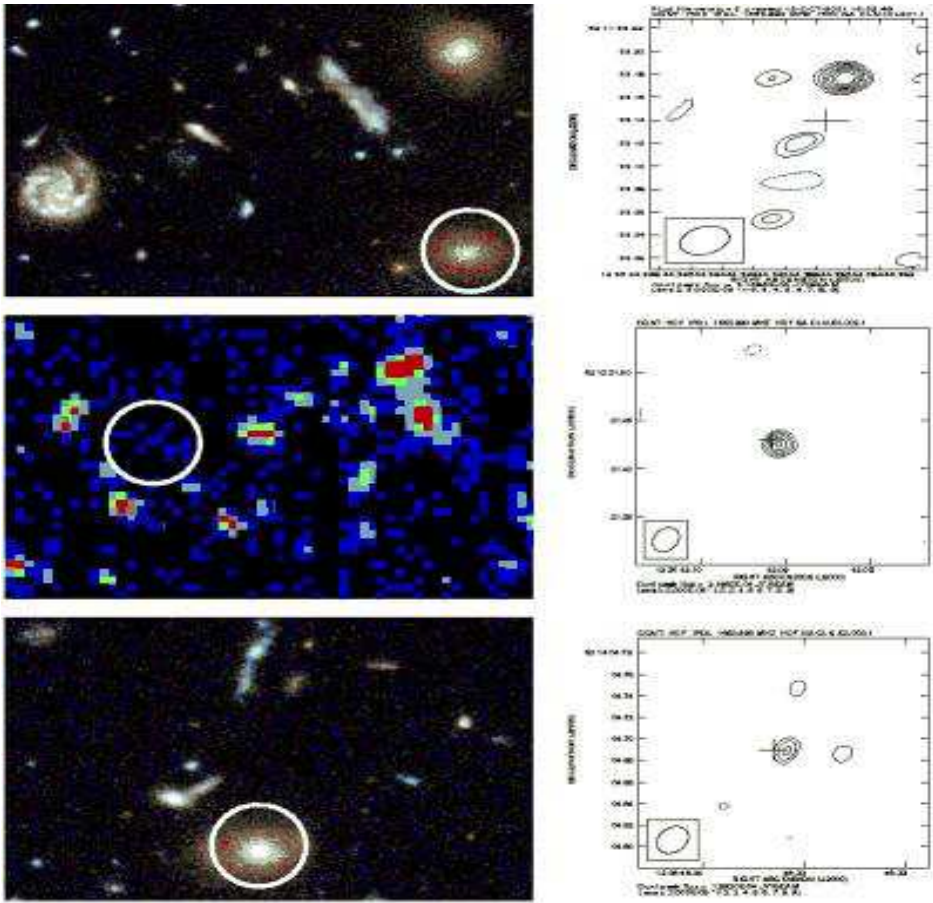}
\vspace{0cm} 
\caption{EVN detections in the HDF: the distant
  z=1.01 FR~I (top), the z=4.4 dusty obscured starburst hosting a
  hidden AGN (middle) and the faint 180 microJy,
  z=0.96 AGN (bottom). Crosses represent the MERLIN-VLA positions for
  these sources.}
\end{figure}

\subsection{EVN Observations of the HDF-N} 

The first deep, wide-field, VLBI observations were conducted by the
European VLBI Network (\cite{gar01}) at 1.6 GHz. Using wide-field,
phase-reference techniques, 3 sources were simultaneously detected
(from a total of 6 potential targets, $S_{1.4 \rm GHz} > 165$~$\mu$Jy)
over a limited radial field-of-view of 3.5 arcminute. Figure 8 shows
the VLBI detections and their optical identification. One of the
sources detected by the EVN (VLA~J123642+621331) is not detected in
I-band ($I > 25^{m}$) and is only just detected in the near-IR in H \&
J-band (\cite{wad99}). Given $K \sim 23^{m}$ (and assuming the K-z
relation holds for these faint systems), Waddington et al. (1999)
identified a single spectral-line detected by the Keck telescope as
Ly~$\alpha$ at a redshift of $z=4.424$.  Although the radio emission is
extended over a scale of 0.6 arcseconds (\cite{mux04}), the detection of
an unresolved radio-loud component by the EVN places a limit on the
brightness temperature of the source, $T_{b} > 10^{5}$~K, strongly
suggestive of an AGN origin for the radio emission. This system is also
detected in the ISO 15$\mu$ supplemental catalogue (\cite{aus99}) and
is an obvious out-lier in the K-corrected and extrapolated FIR-Radio
correlation derived by Garrett (2002). There is little doubt that this
is an optically faint radio source for which the radio emission is very
much dominated by AGN processes.

The ability to use the FIR-radio correlation as a starburst/AGN
diagnostic is likely to improve substantially with the successful
launch and operation of the Spitzer Space Telescope. Already
\cite{app04} have confirmed the FIR/MIR-radio correlation out to $z
\sim 1$ using a matched sample of over 500 sources in the region of the
Spitzer First Look Survey. However, recent studies (\cite{ori04}) show
that about 1/3 of the faint radio source population have no
identification in the Spitzer verification strip at 24 microns. 

\subsection{Deep VLBA-GBT Observations: probing the AGN content of the 
NOAO-N Bo\"otes Deep
  Field.}

Building on the original EVN Deep Field study of the HDF-N, the VLBA
and 100-m GBT telescopes have been used to produce the deepest,
wide-field VLBI images yet - reaching an unprecedented 1 $\sigma$ rms
noise level of $9~\mu$Jy~beam$^{-1}$ (\cite{gar04}). Figure 9 shows the
Bo\"otes field observed by the VLBI array. The survey covered a total
of 1017~arcmin$^2$ $=$ 0.28~deg$^2$ divided into several annular fields.

Nine sources were simultaneously detected out of 61 potential targets
(see Figure 9).  All the sources located in the survey region
have brightness temperatures in excess of $10^{5}$~K, and are likely to
be powered by AGN. For the sample as a whole, the VLBI detection rate
for sub-mJy WSRT radio sources is 8$^{+4}_{-5}$\%.  The VLBI detection
rate for mJy WSRT sources is higher, 29$^{+11}_{-12}$\%. While these
values are lower limits on the AGN content in the field, this trend of
a falling detection rate with decreasing flux density is consistent
with a rapidly evolving radio source population. This {\it direct}
result supports the MERLIN-VLA and WSRT/ISO observations presented in
previous sections, with the sub-mJy and microJy source population
typically being identified with moderate redshift star forming
galaxies.

\begin{figure} % Figure 6
\includegraphics[height=9.0cm,width=12.4cm]{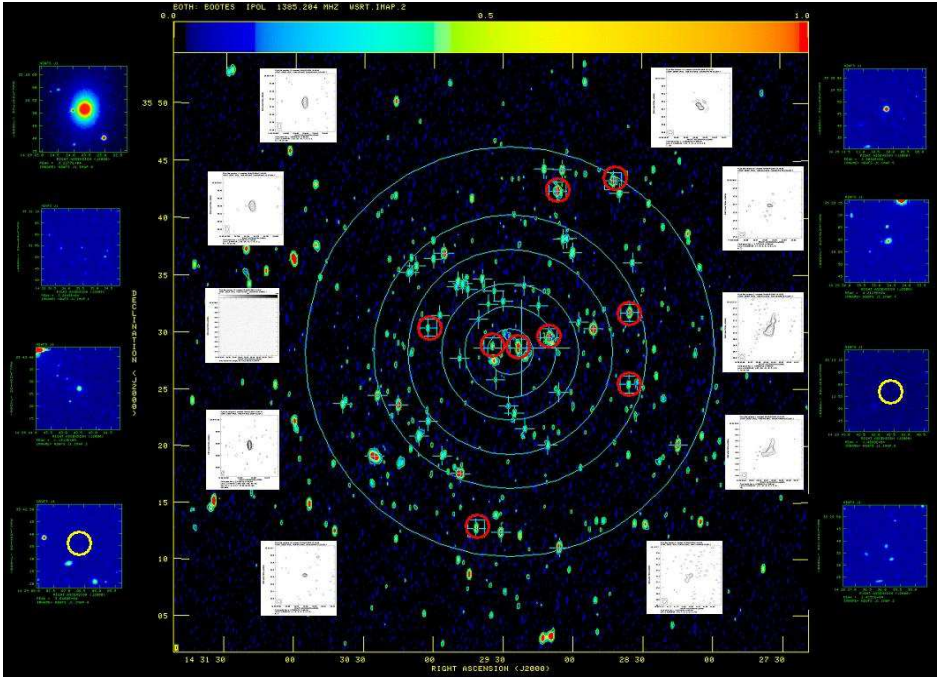}
\vspace{0cm} 
\caption{Nine of the compact radio sources detected simultaneously 
  by VLBI in the Bo\"otes field (contour maps) and (circled) in the
  WSRT finding image. Their optical identifications (surrounding
  images) are also shown and sources with no or only very faint I-band
  detections are circled. }
\end{figure}

Several of the VLBI sources detected in the Bo\"otes field have no or
only very faint optical counter-parts. Limits on their K-band magnitudes 
suggest some significant fraction of these compact radio
sources may also be dust obscured AGN, located at $z > 1$ (similar to
VLA~J123642+621331 in the HDF-N). Figure 9 also presents the I-band
optical images for the VLBI detections \cite{jan04}.

\subsection{Short-term Developments in Deep Field VLBI} 

The simultaneous detection of several sub-mJy and mJy radio sources in
both the EVN and VLBA-GBT deep field observations, suggests that the
{\it combined} response of sources detected within the primary beam of
individual VLBI antennas, may also be used to provide continuous and
accurate phase corrections (via self-calibration of all sources in the
field-of-view) to wide-field VLBI data. Currently this technique of
``full-beam'' VLBI self-calibration is most likely to work well at
1.4~GHz, but with improvements expected at higher frequencies (as
larger recorded bandwidths are employed), the technique might also be
applicable at higher frequencies too.  Certainly at 1.4~GHz, it now
appears that self-calibration ought to be possible towards any random
direction on the sky. Projects such as the PCInt system being developed
at JIVE (\cite{hui04}), will enable the full primary beam to be imaged
out in VLBI observations minimising the losses due to time and
bandwidth smearing.  Wide-field, ``full-beam'' VLBI self-calibration
techniques may also be of benefit to standard, singular faint target
projects. One potential problem is that the computing resources
required for the self-calibration of the huge data sets associated with
wide-field (unaveraged) VLBI data sets is considerable. The use of
cluster (or GRID) computing resources is probably required (Garrett
2004, in prep).

Another important technical development in VLBI is the introduction of
high capacity disk recording systems (\cite{whi03}) and even optical
fibre connections - eVLBI and the e-EVN (e.g. \cite{par04}). Disk
systems are currently deployed across the EVN and should be available
globally in the astronomical community over the course of the next few
years. MicroJy sensitivity levels can be achieved with a global VLBI
array using these new systems, opening up many new possibilities,
particularly in continuum VLBI. 

The surface density of AGN as observed by a high sensitivity,
wide-field VLBI array compare very favourably with other instruments
over equal integration times e.g. Chandra.  Just as there are optically
faint radio sources that are not detected by SCUBA as sub-mm galaxies
(\cite{cha04}) or by Spitzer in the Mid-IR (\cite{ori04}), there are
also a significant number that are not detected by Chandra in X-rays,
even in the deepest 2~Msec HDF-N observations (\cite{bra04}).  As we
are aware in everyday life, hard x-rays are not easily absorbed and
these so-called Compton-thick sources require extremely high column
densities along the line-of-sight towards the obscured AGN ($\sim
10^{24}$~cm$^{-2}$).  It seems then that deep radio surveys can still
be an important source of discovering AGN, that would not otherwise be
detectable at other wavelengths. And high sensitivity VLBI observations
may be an important tool in distinguishing between AGN and the other
``contaminant'' radio source populations {\it e.g.} the numerous star
forming galaxies that dominate the sources detected detected by
lower-resolution radio instruments.

\begin{figure} % Figure 9
\includegraphics[height=11.0cm,width=12.0cm]{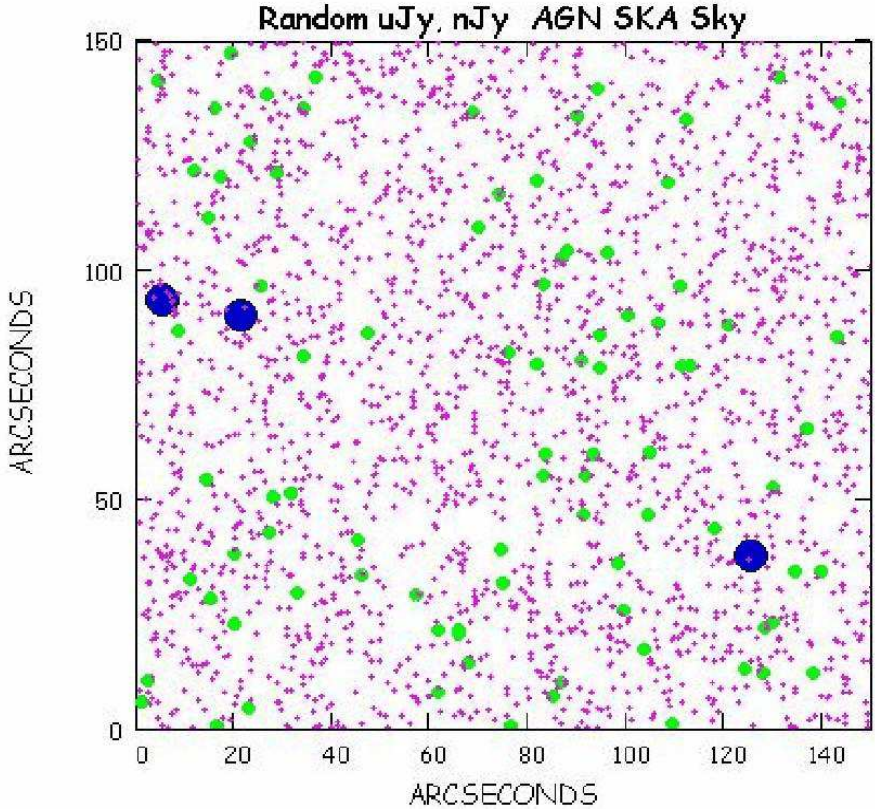}
\vspace{0cm} 
\caption{A random simulation of what the filtered ``AGN sky''
  might look like generally, as seen by the SKA in an integration time
  of 8 hours. The small crosses represent AGN with $S \sim 0.1-1
  \mu$Jy, small filled circles represent AGN with $S \sim 1-100 \mu$Jy,
  and the large filled circles represent AGN with $S > 100 \mu$Jy. The
  extent of the region ($\sim 150$ arcseconds on a side) is similar to
  the original HDF-N.}

\end{figure}

\section{Future Developments: Pin-pointing AGN with the Square Km Array}
\label{SKA}

Recent studies conducted by \cite{hai04}, suggest that a significant
fraction of the faint radio source population may be located at high
redshift. Indeed at the level of 10~$\mu$Jy they predict, in their
simplest models, source surface densities of 10~deg$^{-2}$ at $z > 10$.
The deep field VLBI studies presented earlier, also suggest that at
least some fraction of the compact faint radio sources are located at
$z> 1$. At these sub-mJy and microJy sensitivity levels, VLBI can play
a crucial role in identifying and pin-pointing the location of these
distant radio-loud active galaxies, though it will be difficult to
confirm the redshift of these systems, especially in the case of optically
faint sources.

Next generation radio instruments such as the eVLA/e-MERLIN/eVLBI and
the SKA (Square Km Array) will reach microJy and nanoJy sensitivity
levels respectively. Star forming galaxies are still likely to be an
important component of the source counts at the nanoJy levels probed by
the SKA, but if a significant fraction of the array's collecting area
is located on long baselines ($\sim 4000$ km), it will be possible to
pin-point the location of AGN systems over a field-of-view of at least
1 square degree. In particular, the SKA should be able to distinguish
between star forming galaxies and AGN on the basis of their radio
structure.  For example, star forming galaxies will probably be heavily
resolved, perhaps revealing multiple unresolved components (SNR, SNe)
embedded in extended emission or even more compact clumpy patches of
emission associated with recent merger or nuclear starburst activity.

Figure 10 shows a random simulation of what the filtered ``AGN sky''
might look like generally, as seen by the SKA in a typical integration
time of 8 hours. The extent of the field is similar to the original
HDF-N region. Star forming galaxies are not represented in this plot,
and the limiting sensitivity is $\sim 100$ nanoJy ($> 5\sigma$).  The
simulation assumes (simplistically) that the microJy radio sources
counts can be extrapolated to these sub-microJy flux densities, and
that the fraction of AGN (compared to the total source counts) is the
same as that measured in the Bo\"otes field at sub-mJy levels i.e.
$\sim 8\%$. A more detailed analysis of the fraction of AGN at these
flux density levels is presented by \cite{pra04}. They argue that the
AGN fraction will begin to increase again, at sub-microJy and nanoJy
levels.

Finally, to return to where we started, it's exciting to note, that at
the 100~nanoJy detection levels routinely enjoyed by the SKA, there are over a
100 billion radio sources in the sky!

\begin{acknowledgements} 
  
  I'd like to thank Drs. Loretta Dunne, Raffaella Morganti, Tom Muxlow
  and Joan Wrobel for permitting me to present some of their images and
  results as figures in the text. I'd also like to thank Mrs. Janet
  Eaton for recollecting the exact date of the {\it 57-87} celebrations
  of the Lovell Telescope (Mk1a) at Jodrell Bank. ``I know why you
  can't remember!'' she remarked.

\end{acknowledgements}

\end{document}